\documentclass[prl,lengthcheck]{revtex4-1}
\usepackage{amsmath}
\usepackage{graphicx}
\usepackage{epsfig}
\usepackage{pdfpages}
\def\ket#1{|\,#1 \,\rangle}

\def\Rb{$^{87}\mathrm{Rb}\;$}

\def\uK{\,\mu \mathrm{K}}

\def\nm{\,\mathrm{nm}}
\def\um{\,\mu\mathrm{m}}
\def\mm{\,\mathrm{mm}}

\def\MHz{\,\mathrm{MHz}}
\def\GHz{\,\mathrm{GHz}}

\def\us{\,\mu\mathrm{s}}

\def\E#1{\times 10^{#1}}
\def\eref#1{Eq~\ref{#1}}
\def\fref#1{Figure~\ref{#1}}

\begin{document}
\date{\today}
\author{K. J. Arnold, M. P. Baden, M. D. Barrett}
\affiliation{  Centre for Quantum Technologies and Department of
 Physics, National University of Singapore, 3 Science Drive 2, 117543 Singapore}
\title{Self-Organization Threshold Scaling For Thermal Atoms Coupled to a Cavity}
\begin{abstract}
We make a detailed experimental study of the threshold for self-organization of thermal \Rb atoms coupled to a high finesse cavity over a range of atom numbers and cavity detunings. We investigate the differences between probing with a traveling wave and a retroreflected lattice. These two scenarios lead to qualitatively different behavior in terms of the response of the system as a function of cavity detuning with respect to the probe. In both cases we confirm a $N^{-1}$ scaling of the threshold with atom number.
\end{abstract}
\maketitle
Atoms coupled to the standing wave mode of a cavity will self-organize for sufficiently strong transverse pumping \cite{ritsch2002}. In essence, light scattered into the cavity results in a potential that localizes the atoms to a configuration which favorably enhances collective scattering. Thus above a threshold pump intensity, an initially uniform distribution of atoms will undergo a phase transition, spontaneously reorganizing into a lattice configuration. Self-organization was first observed for thermal atoms in the experiments of \cite{vuletic2003,black2003}, and later with a Bose-Einstein condensate where it was mapped to the Dicke model \cite{esslinger2010,domokos2008}. Self-organization is of particular interest as a platform for cooling as it can be applied to all polarizable particles, including molecules \cite{lu2007,salzburger2009}. In particular, theoretical studies have suggested cooling which has a rate independent of particle number $N$ \cite{ritsch2002,ritsch2011}, in contrast to other ensemble schemes where the cooling rate decreases linearly with $N$ \cite{tajima1998, ritsch2001}. However, numerical simulations suggested that the threshold may scale as $N^{-\frac{1}{2}}$ instead of $N^{-1}$ if statistical fluctuations are required to trigger the self-organization \cite{vukics1}. For large ensembles, an $N^{-\frac{1}{2}}$ threshold scaling places prohibitively severe constraints on the required probe power. Which threshold scaling applies will therefore greatly impact on the viability of self-organization as a cooling method \cite{junye2008}.

\begin{figure}
\centering
\includegraphics[width=10cm]{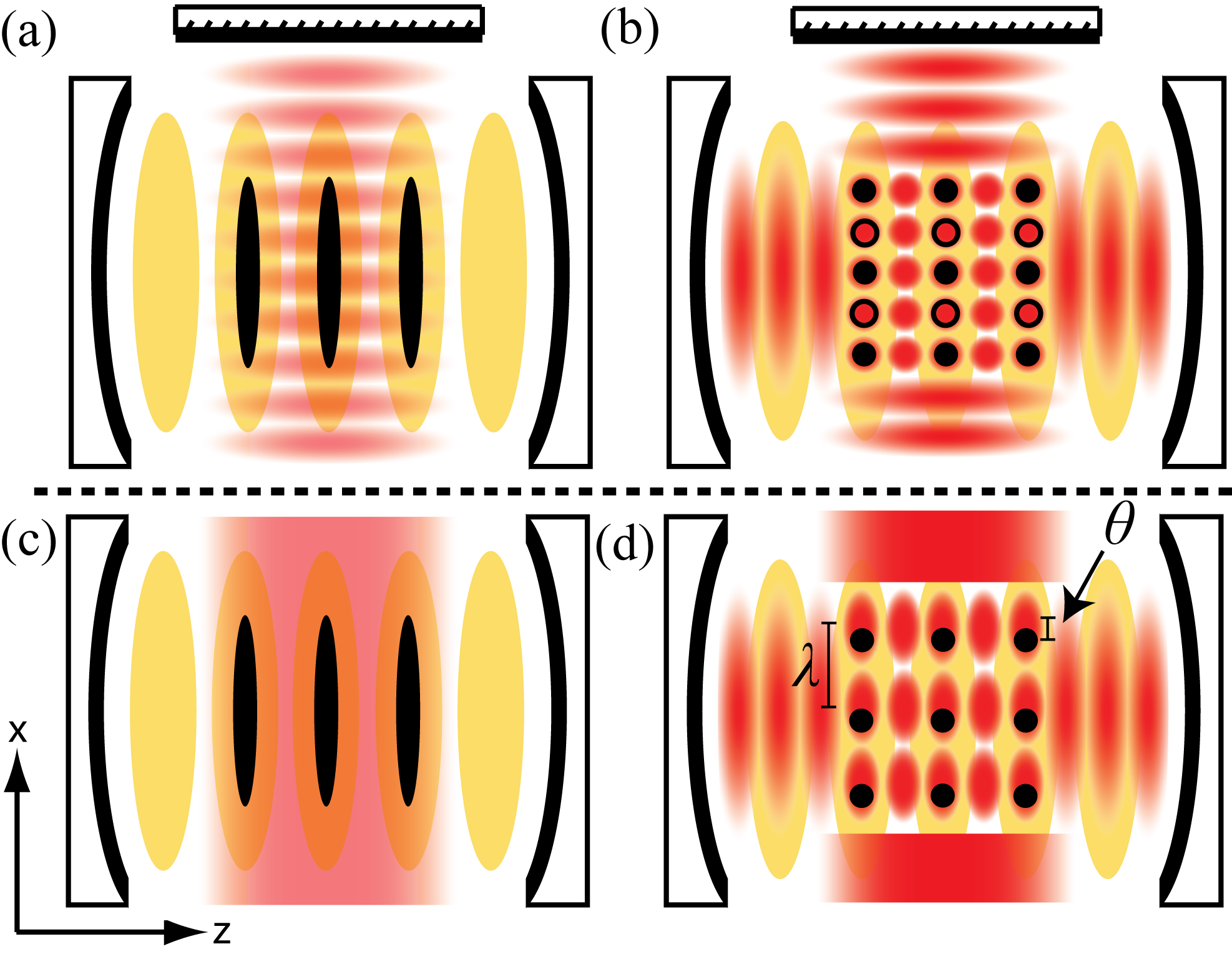}
\caption{Schematic representation of self-organization in the lattice (top) and traveling wave (bottom) geometry. (a, c) Below threshold, the atoms (black) are confined by the intra-cavity 1560 nm FORT (yellow). (b, d) Above threshold, the atoms organize into a $\lambda$-spaced lattice trapped by the probe and scattered fields (red). (b) For a lattice probe, the atoms can form one of two possible $\lambda$-spaced lattices (filled or open circles). (d) For a traveling wave probe, interference between probe and scattered fields results in a $\lambda$-spaced transverse lattice out of phase with the atoms by $\theta$. }
\label{schematic}
\end{figure}

In this Letter, we present a systematic experimental study of the self-organization threshold for $^{87}$Rb atoms trapped in a high finesse optical cavity. We directly measure the threshold behavior over a wide range of experimental parameters for two transverse probing configurations: a retroreflected lattice and a traveling wave, as shown schematically in \fref{schematic}. In both configurations, the atoms are trapped intra-cavity by a $1560\nm$ far-off-resonance optical trap (FORT) which locates the atoms at every second anti-node of the $780\nm$ cavity mode. As discussed in Ref.~\cite{SM}, the threshold behavior depends on both the trapping configuration and the probing geometry used, resulting in a modification to the threshold equation in Ref.~\cite{vukics1}. However, the modified equations for both cases still maintain the $N^{-1}$ scaling within the mean field limit, which our experimental results clearly demonstrate.

The threshold conditions relevant to our system are derived in Ref.~\cite{SM} under the assumption that the system is in thermal equilibrium and that the transverse spatial extent of the atomic ensemble is large relative to the probe wavelength.  For an atom-probe coupling $\Omega$, and detuning $\Delta=\omega_p-\omega_a$ of the probe frequency $\omega_p$ from the atomic resonance $\omega_a$, threshold conditions are conveniently expressed in terms of the dimensionless quantity $\mu=-\hbar \Omega^2/(\Delta k_B T)$, which is the probe trap depth relative to the atoms' thermal energy, $k_B T$. In our system, the threshold equations are given by~\cite{SM}
\begin{equation}
\label{threshold1}
\left(1+\frac{I_1(\mu/2)}{I_0(\mu/2)}\right)\mu =\frac{1}{N U_0 \alpha}\frac{\tilde{\Delta}_c^2+\kappa^2}{\tilde{\Delta}_c}
\end{equation}
for the lattice geometry, and
\begin{equation}
\label{threshold2}
\mu=\frac{\sqrt{\tilde{\Delta}_c^2+\kappa^2}}{-N U_0 \alpha}
\end{equation}
for the traveling wave geometry.  In these equations $I_k(x)$ are modified Bessel functions of the first kind, $\kappa$ is the cavity field decay rate, and $\tilde{\Delta}_c=\Delta_c-N U_0\alpha$. In the last expression $\Delta_c=\omega_p-\omega_c$ is the detuning of the probe frequency from the empty cavity resonance $\omega_c$, and $N U_0 \alpha$ is dispersive shift of the cavity resonance. The dispersive shift is due to $N$ atoms each contributing a maximum single atom dispersive shift $U_0=g^2/\Delta$, where $g$ is the atom-cavity coupling constant. The term $\alpha$ arises from averaging over the atomic spatial distribution. For a $1560\nm$ FORT with antinodes overlapped with every other antinode of the cavity mode, as in our system, $\alpha$ is given by~\cite{SM}
\begin{equation}
\label{alpha}
\alpha=\frac{1}{2}\frac{1+e^{-4/\eta}}{1+2/\eta}
\end{equation}
where $\eta=V_{T0}/(k_B T)$, is the ratio of the trap depth, $V_{T0}$, to the atoms' thermal energy.

The experiments are carried out in a dual-coated high finesse optical cavity. The cavity is $9.6\mm$ long and has a finesse $F=110,000$ near the wavelength of $780\nm$ and $F=160,000$ near $1560\nm$. The high finesse at $1560\nm$ allows us to stabilize the length of the cavity as well as create a deep intra-cavity FORT. The intra-cavity FORT lattice has a waist of $70\um$ and is actively stabilized to a trap depth of 230$\uK$. The $1560\nm$ wavelength allows us to trap the atoms at exactly every second anti-node of the $780\nm$ probe field, such that all trap sites are identically coupled to the cavity mode. The single atom cooperativity is $C=g^2/\kappa\gamma\approx6$ as determined from the cavity QED parameters $(g,\kappa,\gamma) = 2\pi \times (1.1,0.073,3.0)\MHz$, where $g$ is the single atom coupling constant for the $\ket{F=2}$ to $\ket{F'=3}$ cycling transition of the \Rb D2 line, $\kappa$ is the cavity field decay rate, and $\gamma$ is the atomic dipole decay rate~\cite{Steck}.

To load atoms into the cavity, we start from a magneto-optical trap (MOT) $15\mm$ above the cavity. We load up to $8\E{6}$ atoms into a single beam $1064\nm$ wavelength FORT similar to our previous experiments \cite{becpaper,hyperfine}. This FORT beam is moved down $15\mm$ into the cavity over one second by a translation stage. Once in the cavity, the $1064\nm$ FORT is adiabatically ramped off, transferring the atoms into the intra-cavity $1560\nm$ FORT. By varying the MOT atom number, we control the number of atoms delivered to the cavity FORT, up to a maximum of $7\E{5}$.

We probe the atoms with linearly polarized light which is aligned transverse to the cavity axis, as is a magnetic field determining the quantization axis. The experiments were performed at two probe detunings from the D2 transition, $-110\GHz$ and $-265\GHz$. The atoms are optically pumped into the $\ket{F=1}$ ground state manifold resulting in a random distribution of $\ket{m}$ states. However with these detunings and polarization, the distribution across magnetic sub-levels is of no consequence.

\begin{figure}
\includegraphics{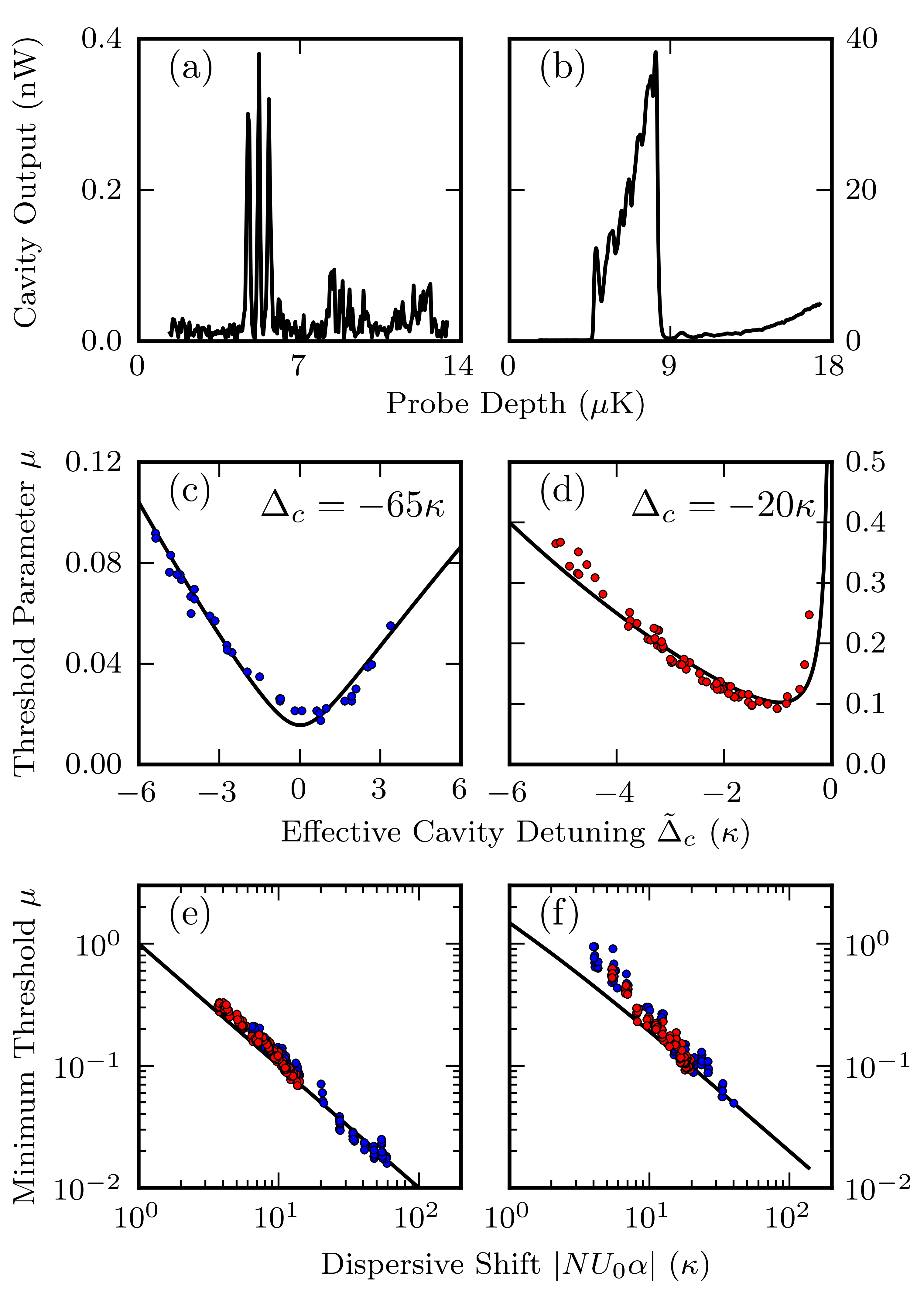}
\caption{ Threshold measurement results for the traveling wave and lattice probe geometries, in left and right columns, respectively. {\bf top:} The self-organization threshold is determined from the sudden increase in cavity transmission as the probe intensity is ramped up over 10 ms, example traces are shown for a traveling wave (a) and lattice (b) probe. {\bf middle:} One set of threshold measurements at fixed $\Delta_c$, for traveling wave (c) and lattice (d) probe. Black lines are an \emph{ab initio} calculation of the threshold from \eref{threshold1} (\eref{threshold2}) for a lattice (traveling wave) probe. {\bf bottom:} All measurements from the data sets at various $\Delta_c$ near the minimum threshold, which for the traveling wave is where $|\tilde{\Delta}_c|\leq\kappa/2$ (e), and for the lattice is where $|\tilde{\Delta}_c+\kappa|\leq\kappa/2$ (f). Black lines are the calculated minimum threshold for the respective cases. In (c-f) red circles are data taken at the atomic detuning $-265\GHz$, and blue circles at $-110\GHz$.}
\label{masterfig}
\end{figure}

To verify the threshold equations, we need to measure the dispersive shift and the threshold intensity. The dispersive shift is non-destructively measured by sweeping the frequency of a weak probe beam coupled to the cavity over the cavity resonance and observing the cavity transmission. Operating at a known laser-cavity detuning, $\Delta_c$, we are able to infer the dispersively shifted detuning, $\tilde{\Delta}_c$, from the directly measured dispersive shift, $N U_0 \alpha$. Next the threshold is measured by linearly ramping up the intensity of a transverse probe beam over 10 ms while monitoring the cavity output. The cavity output is fiber coupled and split between a single photon counting module (SPCM) and heterodyne detection setup. The SPCM is used for detecting weak signals, while the heterodyne detection is used to detect signals which would otherwise saturate the SPCM. The threshold for self-organization is clearly observed as a sudden increase in cavity output, as illustrated in \fref{masterfig}(a).

In order to properly compare the measured threshold power to the threshold equations three effects must be accounted for. First, the cavity has a significant birefringence separating the two linear polarization modes by $5\kappa$. Limited optical access constrains the probe beam polarization to be misaligned by 21$^\circ$ with respect to the cavity mode polarization. This reduces the scattering rate into the cavity by 13$\%$. Second, in the lattice configuration, ramping up the probe beam results in a significant temperature increase. Since this temperature increase is not observed for the traveling wave probe, it is most likely due to adiabatic compression. As the threshold parameter $\mu$ depends on the temperature at threshold, we must calibrate this temperature increase. This is done by measuring the temperature throughout an adiabatic ramp of the probe power with the cavity far detuned from the probe in order to avoid the onset of self-organization. Third, the dispersive shift also depends on temperature via the parameter $\alpha$. Thus, we compensate the measured dispersive shift using \eref{alpha} taking into account the initial temperature, $31\pm2\uK$, and the temperature at threshold as determined from the temperature calibration measurements.

For the traveling wave geometry, \fref{masterfig}(c) shows the results of several threshold measurements at fixed probe detuning, $\Delta_c$. By coarsely controlling the mean dispersive shift via the atom number, we sample a range of $\tilde{\Delta}_c$ such as the data set shown. The threshold parameter is obtained by scaling the probe depth at which light is scattering into the cavity by the temperature at threshold. The black line shows the theoretical threshold calculated \emph{ab initio} from \eref{threshold2}. Data sets such as \fref{masterfig}(c) were taken at several values of $\Delta_c$ ranging from $-65\kappa$ to $-5\kappa$. In \fref{masterfig}(e) the black line shows the minimum threshold calculated from \eref{threshold2} at $\tilde{\Delta}_c=0$ and all measurements near the minimum for which $|\tilde{\Delta}_c|<\kappa/2$. As can be seen from the figures, the threshold measurements are in good agreement with Eq.~\ref{threshold2} and a $N^{-1}$ scaling.

For the lattice geometry, the probe beam is retro-reflected but we otherwise measure the threshold in the same way as for the traveling wave case. \fref{masterfig}(d) shows an example data set at fixed $\Delta_c$. The black lines are calculated from \eref{threshold1}, where the minimum threshold occurs at $\tilde{\Delta}_c=-\kappa$, in contrast to the traveling wave case. \fref{masterfig}(f) shows all measurements near the minimum threshold, for which $|\tilde{\Delta}_c+\kappa|<\kappa/2$. Again the results are in reasonable agreement with Eq~\ref{threshold1} and a $N^{-1}$ scaling.

However, in the lattice configuration there is an increasing discrepancy as the threshold parameter $\mu$ increases. We believe this to be a consequence of our trapping geometry. Because the $1560\nm$ FORT restricts the axial motion of the atoms, the probe lattice itself acts as a potential barrier the atoms must overcome in order to organize. Clearly, in the limit that $\mu \gg 1$ and the axial confinement is large, the atoms will not be able to organize at all. However, we observe a significant slow down in the transition to the organized phase near threshold even for $\mu$ approaching unity. We suspect the organization is initially limited by the time required for the scattered potential to reduce the barrier induced by the probe and facilitate runaway self-organization. This is supported by the observation that organization again becomes rapid if we probe well above threshold. These effects result in the deviation of the measured threshold in \fref{masterfig}(f) as $\mu$ approaches unity. When we ramp the probe power, the slow onset of self-organization near to the threshold results in a delay between crossing threshold and detecting a significant cavity output. Thus the threshold power is systemically overestimated. If the atoms were not confined in the axial direction, we would expect rapid self-organization to occur for $\mu \geq 1$. The absence of any systematic deviation in the traveling wave case, as seen in \fref{masterfig}(e), is consistent with this expectation. In this case there is no lattice potential prior to organization so the atoms are not confined in the transverse direction.

\begin{figure}
\includegraphics{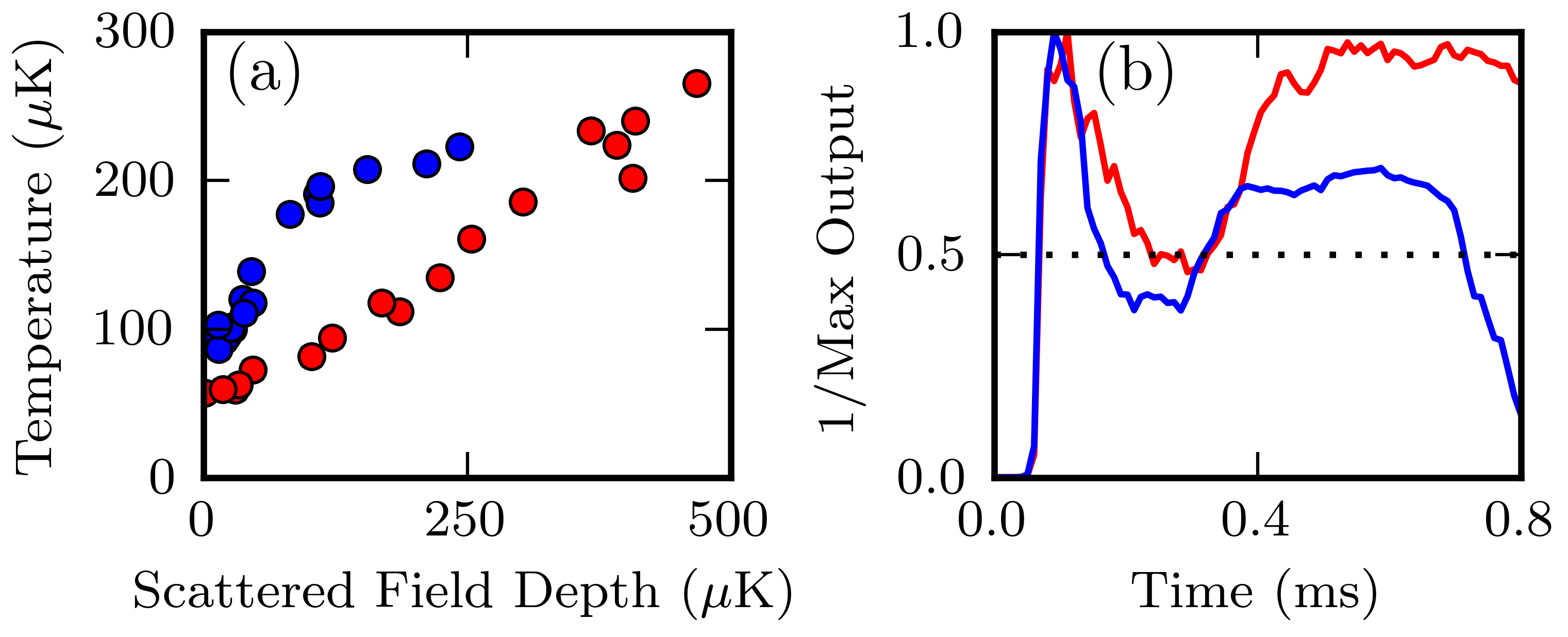}
\caption{Indication of excess heating due to non-adiabatic dynamics after self-organization. (a) Measured temperature after self-organization by probing above threshold at constant power for 0.8 ms plotted against the depth of the scattered field in the cavity as determined by the cavity output at the time of measurement. (b) Two example traces of the cavity output scaled the peak output corresponding to data points in (a). For red line (circles) the cavity output was above 50$\%$ of its peak at the measurement time, whereas for blue line (circles) the transmission was already below 50$\%$ and falling. For reference, an intra-cavity photon number of $4\E{6}$ corresponds to a $530\uK$ deep optical lattice potential ($\Delta=-265\GHz$).}
\label{temperature}
\end{figure}

After the onset of self-organization, the cavity transmission traces, for example \fref{masterfig}(a-b), indicate complex dynamics for both probe geometries. In the case of the lattice probe configuration, our external potential geometry restricts us to $\mu < 1$ for the reasons discussed above, and therefore we must operate in the regime of large dispersive shift ($|N U_0 \alpha| \gg \kappa$). In this regime a small change in $N U_0 \alpha$ results in a significant change in $\tilde{\Delta}_c$. This results in non-linear dynamics above threshold due to the strong inter-dependence of the $T$, $\alpha$, $\tilde{\Delta}_c$, and $\mu$. We observe the organized phase persists for at most several milliseconds, as it does in \fref{masterfig}(b). We suspect an increasing temperature results in the system falling below threshold and reverting to the unorganized phase. This is supported by the data shown in \fref{temperature}. Here the lattice was held at constant power above threshold and the temperature was measured via ballistic expansion after probing for 0.8~ms. The hysteresis in the temperature as the scattered potential rises and falls indicates that the temperature rise is not simply due to adiabatic compression. It suggests the presence of a strong heating mechanism possibly a result of non-adiabatic dynamics induced by the rapidly changing potential.

In the case of the traveling wave, the dynamics above threshold are complicated by the fact that the induced organizing potential is out of phase with atoms' positions by $\theta=\tan^{-1}(-\kappa/\tilde{\Delta}_c)$, as discussed in Ref.~\cite{SM}. When $|\tilde{\Delta}_c| \lesssim \kappa$, this results in the organization switching off as the atoms are pushed away from the axial center of the cavity mode, decreasing the coupling to cavity and increasing the threshold condition. This effect can be seen from the cavity transmission trace \fref{masterfig}(a). Presumably as the atoms move back into the cavity mode, self-organization is re-initiated, resulting in the observed pulsing output. In principle the lattice can be stabilized with an additional force, as demonstrated for the CARL~\cite{kruse2003}. In our case, a displacement in the transverse Gaussian potential can counteract the phase and form a stable superradiant configuration. When $|\tilde{\Delta}_c| \gg \kappa$,  only small displacement is required since $\theta\approx0$, and we indeed observe stable output for several milliseconds.

In summary, we have shown that the self-organization threshold agrees well with a model based on simple mean field considerations given in Ref.~\cite{SM}. Earlier numerical work using multiparticle simulations indicated that scaling strongly depends on model details and assumptions~\cite{vukics1}. However we would argue that the mean field approach should yield accurate results for the threshold, provided there are sufficiently many atoms and a physical mechanism by which the atoms thermalize on the timescale of interest. In our experiments, the densities are such that collision times are $\sim100\us$ with particle numbers of $\sim10^4$ per site of the $1560\nm$ lattice potential. In this case we can expect a thermodynamic description to be valid, and, indeed, the mean field results apply as evidenced by our measurements. For cases in which a thermodynamic description may not apply, such as for fermions or low densities of molecules where the collision rate is negligible, a sub-$N^{-1}$ scaling of the effective threshold may still be applicable.

Our experiments also indicated a strong heating mechanism above threshold contrary to the cooling seen in the numerical simulations ~\cite{vukics1} but consistent with statements made in Ref.~\cite{ritsch2011}. In our case the heating is most likely due to the complex dynamics involved as the atoms organize. In future experiments, we will use a transverse $1560\nm$ lattice to confine the atoms. This will enable us to work more effectively in the regime of small dispersive shifts and to explore recent theoretical proposals for dissipation induced self-organization~\cite{ritsch2011}. Additionally we will be able to establish a two dimensional $\lambda$-lattice. In this configuration, the atoms are expected to scatter super-radiantly without threshold into the cavity mode. In this context we can study potential cavity cooling mechanisms using weaker pumping and without the complex dynamics of the optomechanical forces driving the self-organized phase.

We acknowledge the support of this work by the National Research Foundation and the Ministry of Education of Singapore, as well as by A-STAR under Project No. SERC 052 123 0088.


\end{document}